# Bistability of Free Cobalt and Iron Clusters

Xiaoshan Xu, Shuangye Yin, Ramiro Moro, Anthony Liang,

John Bowlan, Walt A. de Heer

School of Physics, Georgia Institute of Technology, Atlanta GA

**Abstract** The cobalt and iron clusters $Co_N$, $Fe_N$ ($20 < N < 150$) measured in a cryogenic molecular beam are found to be bistable with magnetic moments per atom both $\mu_N/N \sim 2\mu_B$ in the ground states and $\mu_N^*/N \sim \mu_B$ in the metastable excited states (for iron clusters, $\mu_N \sim 3N\mu_B$ and $\mu_N^* \sim N\mu_B$). This energy gap between the two states vanish for large clusters, which explains the rapid convergence of the magnetic moments to the bulk value and suggests that ground state for the bulk involves a superposition of the two, in line with the fluctuating local orders in the bulk itinerant ferromagnetism.

Ferromagnetism of bulk crystalline iron and cobalt are often explained by electronic band theory as the consequence of global population imbalance of partly filled electronic spin-up and spin-down bands of itinerant electrons [1]. To account for magnetic properties at elevated temperatures, small fluctuating domains with local orders have to be accommodated [2, 3]. In fact, the itinerant or localized natures of metallic ferromagnetism have been intensely debated for over a half century starting with pioneers of the quantum theory of solids coinciding with the development of band theory [1, 4, 5]. Despite the great deal of efforts to solving this problem, to date, ferromagnetism of iron and cobalt is still not fully understood. On the other hand, the knowledge of the atomic magnetism is thought to be much more comprehensive. Therefore, the magnetism of metal clusters is of great interest not only because it demonstrates the evolution of magnetism from isolated atoms to the bulk, but also because it may be directly related to the proposed small domain of local orders in the bulk.

Previous measurements show that the magnetic moments of small cobalt clusters and iron clusters are enhanced: for $Co_N$, $\mu_N/N\mu_B \approx 2$ while for $Fe_N$, $\mu_N/N\mu_B \approx 3$. For larger clusters ($N>700$) these values converge to their respective bulk values [6-8]. Since surface atom spins are expected to be more localized it was originally conjectured that this convergence reflected the diminishing role of the surface [9-14]. However, the convergence is too rapid to be explained by geometry [6, 13, 14] (for a cluster of 700 atoms, more than 40% of the atoms are on the surface, however the moments already converge to the bulk value).

We show here that small clusters are bistable with ground states (GS) and metastable excited states (MES). For both cobalt and iron clusters, the magnetic moments $\mu_N^*/N$ of the MES are approximately 1 $\mu_B$, which is lower than the bulk value, indicating localized moments. A population analysis of the cluster beam indicates that the GS and the MES become degenerate for large size, while ionization potential measurements indicate that the energy gap between these two states closes with increasing size. This may account for the rapid convergence of the magnetic moments to the bulk value. In addition, this may be related to the fluctuating domains with local orders in the bulk itinerant ferromagnetism, suggesting that in clusters the itinerant ferromagnetic state evolves from two states with different chemical valences.

Cluster magnetic moments $\mu_N$ can be determined by deflecting cold cluster beams (20 K $\leq T \leq$ 100 K) in an inhomogeneous magnetic field $B$ [6, 15-17]. The magnetization $M$ of a specific cluster is the average projection of its magnetic moment along the magnetic field direction which depends on the state of the cluster. The deflection $\delta$ of a cluster is linearly proportional to its magnetization: $\delta = K(dB/dz)/(mv^2)M$, where m is the mass of the cluster, $v$ is its speed and $K$ is a constant that depends on the geometry of the apparatus. The magnetization distribution $P(M)$ of an ensemble of clusters of a given size is determined from the shape of the deflected beam. The average magnetization $<M>$ of this ensemble produced in a cluster source at temperature $T$ approximately follows the Langevin function $L$: $<M>=L(\mu B/k_B T)$ where $k_B$ is the Boltzmann constant [17-19]. Hence for each cluster size $N$, $\mu_N$ can be extracted from $<M>_N$.

In brief, the experimental methods are as follows (for details see Ref 6, 15-17, 20). A YAG laser vaporizes a small amount of metal from the sample rod located in the source. Simultaneously, a pulse of cryogenically cooled helium gas is injected into the source. The metal vapor is cooled and condenses into clusters. The clusters dwell in the cold source for about 1ms, after which they exit the nozzle into the vacuum chamber resulting in a beam of clusters. The cluster beam is collimated by 0.1mm x 5mm slits. After traveling for about 1 m in high vacuum the beam passes between the pole faces of a Stern-Gerlach magnet that causes the magnetic clusters to deflect. The neutral clusters enter the detector chamber where they are photo-ionized with light from a tunable pulsed laser. Positions and masses of the deflected clusters are measured with a position-sensitive time-of-flight mass-spectrometer located at the end of the beam, about 2 m from the source. Beam speeds are measured using a mechanical beam chopper. Ionization efficiencies (IEs) are determined by recording the cluster ion intensities as a function of the ionization photon energy. The ionization potentials (IPs) are determined from the IEs. Cluster polarizabilities are determined from the cluster deflections in an inhomogeneous electric field.

The magnetization probability distribution profiles $P(M)$ for cobalt and iron clusters are shown in Fig. 1. The magnetic moments found from Fig. 1(a) and Fig. 1(d) are consistent with previous measurements but with better precision [8].

It is known that magnetic deflections are sensitive to source conditions and that metastable states can survive under certain circumstances [21]. Here we found that reducing the amount of cold helium gas injected into the source causes a distinct double peaked structure [Fig. 1(b)]. The deflections of the first peak were identical to those in Fig. 1(a). The second peaks correspond to MES $Co_N^*$. We carefully ruled out any artifacts specifically those involving bimodal operation of the source, in fact the effect was observed in three different sources. Moreover others may have observed the effect [7]. The speed distribution of the MES clusters is identical to that of the GS clusters, indicating that the translational temperature (and the rotational temperature) is equilibrated with the source [22]. For $Co_N$ we find that the magnetic moment and hence the spin of the MES is about half of that of the ground state. The magnetic moments themselves are found not to depend on source conditions providing further confidence in the interpretation. The MES appear for all cluster sizes, temperatures and magnetic field ranges ($20 \leq N \leq 200$, $20K \leq T \leq 100K$, $0T \leq B \leq 2T$) in our experiments. Figure 1(d-f) shows the same effect for iron clusters.

The following experiment confirms that the low magnetic moment states are indeed metastable. The cobalt cluster beam was illuminated with a pulse of 500 nm laser light before it entered the magnetic field. The *P(M)* are compared with those without laser without laser heating. As can be seen in Fig. 2(a-c), intensity in *P(M)* is transferred from peak to the $Co_N^*$ peak. Apparently heating by one (or a few) photons converts a fraction of the clusters from the GS to the MES.

The IPs of the two states were derived from their ionization efficiencies. These were determined by recording the cluster ion intensity as a function of laser wavelength from 250 nm to 215 nm while the source conditions were tuned to either of the two states. Figure 2(d) shows the IPs for $Co_N$ and $Co_N^*$. The IP difference between the GS and the MES are on the order of 0.1 eV for small clusters and vanish for larger clusters ($N>150$). Note that for Co$_N$, the $IP_N$ are particularly high for $N=34$ and $N=40$ which suggests electronic shell effects.

For GS cobalt clusters $\mu_N/N$ decreases slightly with increasing $N$ and converges to about 2 $\mu_B$ at $N\sim150$ [Fig. 3(a)]. For Fe$_N$, $\mu_N/N$ is close to 3 $\mu_B$ [Fig. 3(b)] for all $N$. The magnetic moments $\mu_N^*/N$ of MES cobalt and iron clusters converge to 1$\mu_B$ for both Co$_N^*$ and Fe$_N^*$ (Fig. 3).

The electric dipole polarizabilities of cobalt and iron clusters are measured by deflecting the cluster beam in an inhomogeneous electric field [16]. For cobalt clusters [Fig. 4(a)] the GS polarizabilities $\alpha_N$ have larger values than the MES polarizabilities $\alpha_N^*$. The GS clusters show remarkable large undulations whereas the MES polarizabilities decrease monotonically. For iron clusters, $\alpha_N$ and $\alpha_N^*$ are similar [Fig. 4(b)].

The polarizability of a classical metal sphere is $R_N^3$ where $R_N = R_1 N^{1/3}$ is the classical cluster radius. The electronic spillout effect enhances the polarizablity: $\alpha_N=(R_N+d)^3$ where $d$ is of the order of 1 Å [20]. The 4s electrons are more delocalized than the 3d electrons so that they spill out more than the 3d electrons. Consequently, they are primarily responsible for the enhanced polarizabilities and for shell structure effects. Hence, the structure in the polarizabilities of Co$_N$ and the absence of structure in Co$_N^*$ suggests that Co$_N$ clusters have 4s electrons while Co$_N^*$ do not. This is also consistent with the shell structure in the IP measurements of Co$_N$. In contrast, the polarizabilities of Fe$_N$ and Fe$_N^*$ are rather similar and featureless [Fig. 4(b)].

Summarizing the experimental observations, cobalt and iron clusters are bistable with two distinct magnetic states: a high moment GS (for Co$_N$, $\mu_N/N\mu_B \approx 2$; for Fe$_N$, $\mu_N/N\mu_B \approx 3$) and a low moment MES ($\mu^*_N/N\mu_B \approx 1$). For Co$_N$, the IP and the polarizability measurements indicate that the atomic electronic configurations of the states are different: the GS appear to have 4s electrons (giving rise to enhanced polarizabilities and structure in the IPs) which are absent in the MES. The energy difference between the MES and GS diminishes with increasing size and for large sizes the two tend to produce a mixed ground state with bulk-like fractional magnetic moments as previously observed.

Next, we try to understand the nature of the MES and the bistability of the Co and Fe clusters according to their measured electric and magnetic properties. Although this bistability has not been observed in supported nanostructures, for example, magnetic atomic chains, magnetic thin films, the supported nanostructures and free clusters do share common grounds such as reduced atomic coordinations and reduced spatial electronic densities, which both decrease the interatomic overlap of electronic wave functions and in turn make the electrons more localized and enhance the magnetic moments [6-8, 23]. In addition, absence of translational symmetry of free clusters makes the electrons even more localized [24], particularly for 3d electrons, whose band width is already narrow in the bulk limit. In contrast, 4s electrons are always delocalized and with large band width. In this context, we may discuss the 3d electrons of Co and Fe as if they were localized on atoms and obeyed Hund's rules, keeping in mind that the orbital angular moments are quenched by the loss of spherical symmetry of atomic centers, i.e. the magnetic moments are from electronic spins. The assumption here is not the isolation of individual atoms, but the tendency to more Heisenberg-like [5] rather than Stoner-like [1] exchange interactions for small clusters because of the reduced dimensionality and the spatial electronic densities. Following this assumption is the integer number of magnetic moments per atom in $\mu_B$, which appears to be consistent with our observations. The magnetic moments suggests the atomic spin state $S=3/2$ for $Fe_N$ and $S^*=1/2$ for $Fe_N^*$. For cobalt clusters, $S=1$ and $S^*=1/2$.

Hence, the approximate electronic configurations of the states can be derived with some confidence, again, assuming the localized 3d electrons. The electronic configuration of the free cobalt atom is $3d\uparrow^5 3d\uparrow 4s^2$. The electronic structure of ground state $Co_N$ clusters appears to be primarily derived from atomic orbitals with the configuration $3d\uparrow^5 3d\downarrow^3 4s^1$ [25] to produce the observed spin magnetic moment $\mu_N/N\mu_B \approx 2$. Similarly, the atomic configuration of $Co_N^*$ appears to be dominated by $3d\uparrow^5 3d\downarrow^4 4s^0$ orbitals so that $\mu_N/N\mu_B \approx 1$ as observed. It is satisfying to note the presence of 4s electrons in the ground state, and their absence in the MES, agrees with the polarizability and IPs observations above. For $Fe_N$, the configuration is $3d\uparrow^5 3d\downarrow^2 4s^1$ to produce the observed $\mu_N/N\mu_B \approx 3$, while for $Fe_N^*$ we assume that it is $3d\uparrow^4 3d\downarrow^3 4s^1$ so that $\mu_N/N\mu_B \approx 1$. The 4s character is similar in the GS and MES which explains the similar polarizabilities. The electronic configuration disparity of these two states in both Co and Fe clusters inhibits decay [22] from one to the other, which explains the metastability of the excited state.

The question remains why there are two distinct states with significantly different electronic properties in all of the clusters and how these states converge at large size. The bistability of Co and Fe clusters are not described in the literature [10-14] although it may be related to the small domains with fluctuating local order observed in the bulk [2,3]. In fact, the spin weighed average magnetic moments per atom over the GS and MES are 1.7$\mu_B$ and 2.3 $\mu_B$, for iron and cobalt clusters respectively, which are very close to the bulk values (1.7$\mu_B$ for cobalt and 2.2$\mu_B$ for iron). We next provide a tentative description in terms of the interaction between the 3d localized electrons and 4s itinerant electrons. Following the Falicov-Kimball model [26], the Hamiltonian of an $N$ atom cobalt cluster is

$$H = \sum_{k,\sigma} \varepsilon_k C^+_{k\sigma} C_{k\sigma} + \sum_{i,\sigma} \varepsilon_d d^+_{i\sigma} d_{i\sigma} + G \sum_{i,\sigma,\sigma'} C^+_{i\sigma} C_{i\sigma} d^+_{i\sigma'} d_{i\sigma'},$$

where $C_{k\sigma}^+$ ( $C_{k\sigma}^+$) destroys (creates) an electron in 4s band, $d_{i\sigma}^+$ ( $d_{i\sigma}^+$) destroys (creates) an electron in the 3d Wanier state, $C_{i\sigma}^+$ ( $C_{i\sigma}^+$) destroys (creates) an electron in 4s Wanier state, $\varepsilon_k$ ($\varepsilon_d$) is the energy of 4s (3d) electrons and $G$ is the intra-atomic repulsion. Hence the Hamiltonian describes the interchange of 4s and 3d electrons for fixed $N=N_s+N_d$, where $N_s$ ($N_d$) is the number of the 4s (3d) electron. The total energy has minima $N_s=N$ and $N_s=0$ with energies $E_1=W/2N+\varepsilon_k N$ and $E_2=\varepsilon_d N$ respectively, where $\varepsilon_k$ is the bottom of and $W$ is the width of the 4s band respectively. For small clusters, $W/2<\varepsilon_d-\varepsilon_k$ so that the two minima represent a GS of configuration $3d\uparrow^5 3d\downarrow^3 4s^1$ with energy $E_1$ and MES of configuration $3d\uparrow^5 3d\downarrow^4 4s^0$ with energy $E_2$, separated by a barrier of height $(G-\varepsilon_d+\varepsilon_k)^2/(4G-2W)+N\varepsilon_d$ at $3d\uparrow^5 3d\downarrow^{3+x} 4s^{1-x}$ where $x = 1/2+(W-2\varepsilon_d+2\varepsilon_s)/(4G-2W)$. Hence this model yields the two states as well as the metastability of the excited state. The 4s electron density increases with increasing cluster size because the ratio of the volume occupied by a 4s electrons in a cluster to that in the bulk is approximately $(1+d/R_N)^3$ [20]. This causes both $G$ and $W$ to increase so that for sufficiently large $N$. For large sizes, the energy difference between the two states $E_2-E_1=(\varepsilon_d-\varepsilon_s-W/2)N$ vanishes and the two states become degenerate as experimentally observed.

To conclude, we have experimentally demonstrated that both iron and cobalt clusters are bistable with GS and MES whose magnetic moments are higher and lower than the bulk value. The two states become degenerate for sufficiently large clusters, which explains the rapid convergence of the magnetic moments to the bulk value. In addition, the bistability of the Co and Fe clusters may be associated with the fluctuating domains of local orders in the bulk and the evolution from localized magnetism in small clusters to bulk-like itinerant magnetism as observed in large clusters.

**Captions**

FIG. 1 (color). Probability distributions of magnetizations $P(M)$ for cobalt and iron clusters for various thermalization conditions. Amplitudes are represented in color (blue:low; red:high). (a-c) $P(M)$ for cobalt clusters of 20-200 atoms under good, intermediate and restricted thermalization conditions at $T$=20K, $B$=2T. The two branches correspond to ground state $Co_N$ and metastable state $Co_N^*$ clusters. The proportion of $Co_N$ and $Co_N^*$ can be tuned continuously, but their magnetic moments are not affected. (d-f) $P(M)$ for iron clusters containing 20-200 atoms under good, intermediate and restricted thermalization conditions at $T$=20K, $B$=1.2T, revealing two states: $Fe_N$ with about 3 $\mu_B$ per atom and $Fe_N^*$ with 1$\mu_B$ per atom.

FIG. 2 (color). (a) Probability distributions of magnetization $P(M)$ for cobalt cluster of 30 atoms at 30 K. Note that when the cluster beam is heated by a 500 nm laser before it enters the magnetic field (dashed line), some of the $Co_{30}$ are converted into $Co_{30}^*$. $P(M)$ of cobalt clusters of 10 to 100 atoms without (b) and with (c) laser heating are also shown. Amplitudes are represented in color (blue:low; red:high). Laser heating has the same effect as restricted thermalization. (d) Ionization potentials of $Co_N$ ($IP_N$) and $Co_N^*$ ($IP_N^*$).

FIG. 3. Magnetic moments per atom for cobalt (a) and iron (b) clusters. The magnetic moments are deduced from low field data for which in general $<M>=\mu B^2/3k_BT$.

FIG. 4. Electric dipole polarizabilities of cobalt (a) and iron (b) clusters.

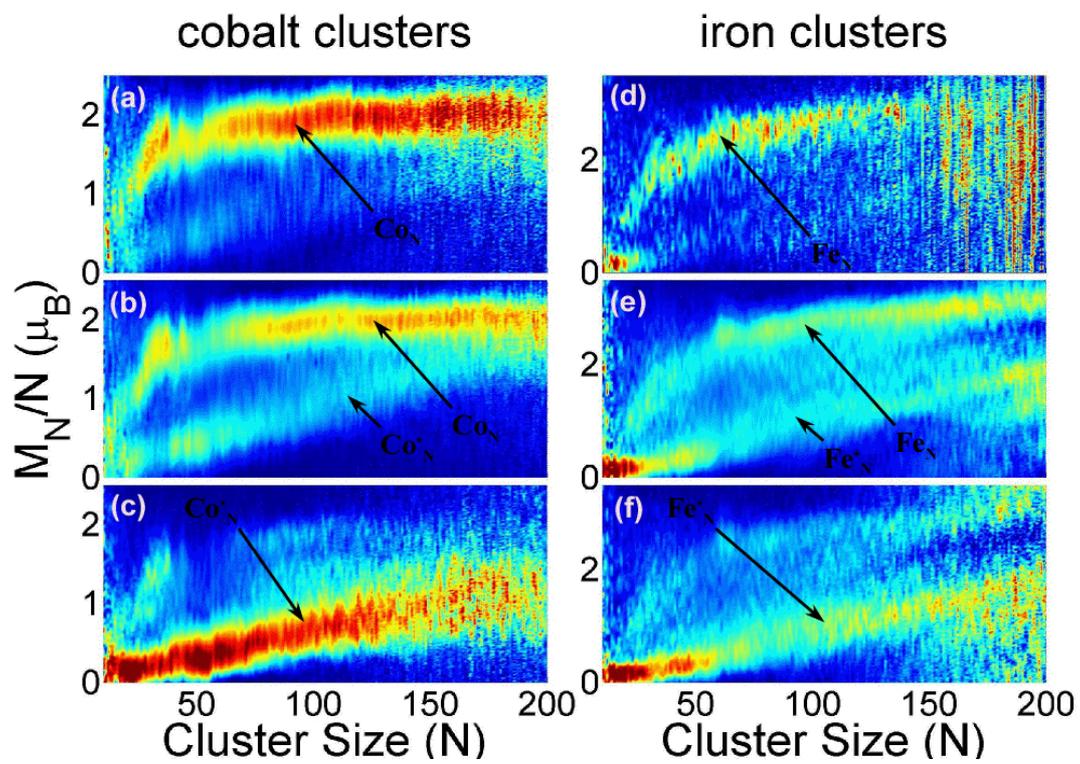

Figure 1

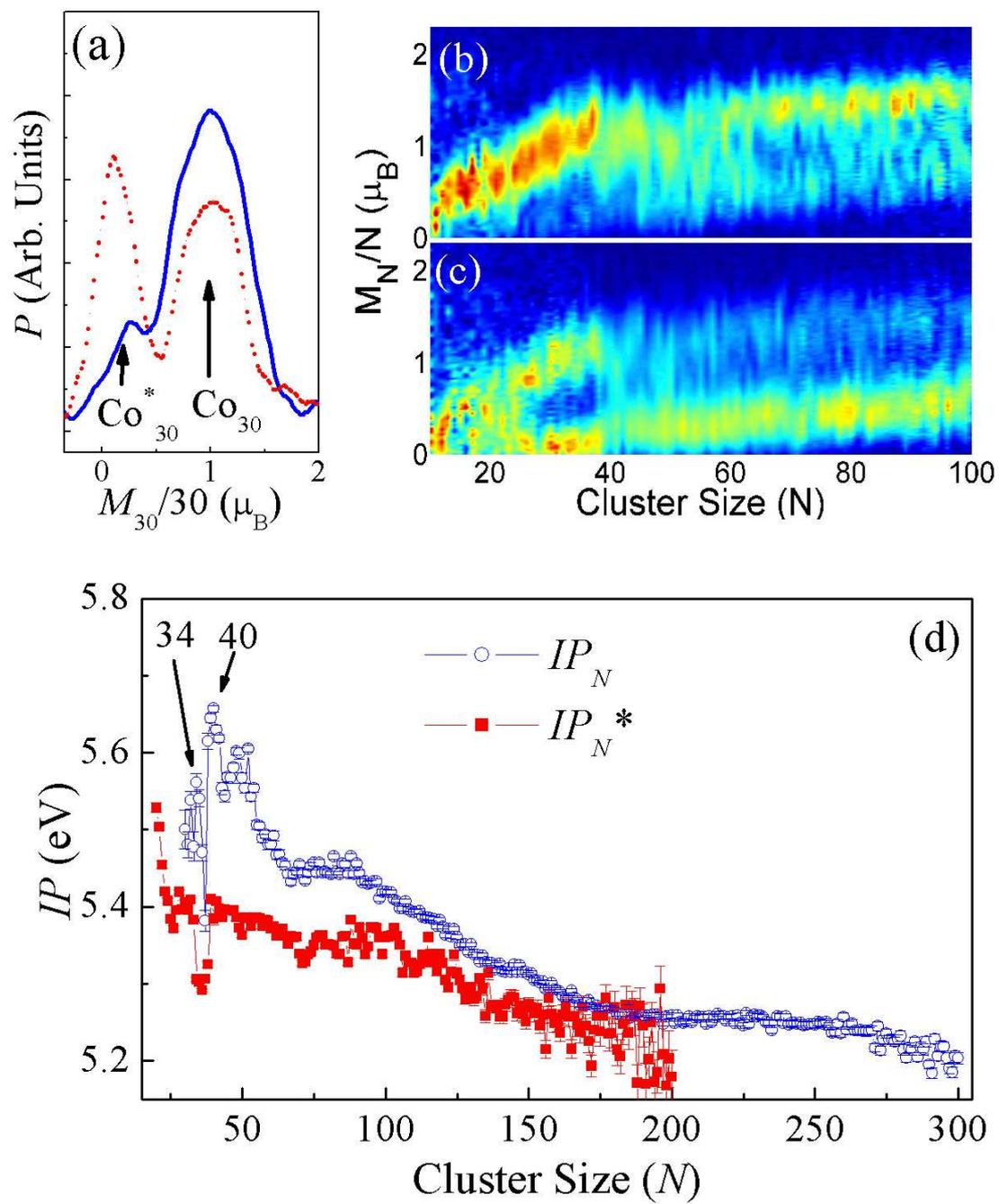

Figure 2

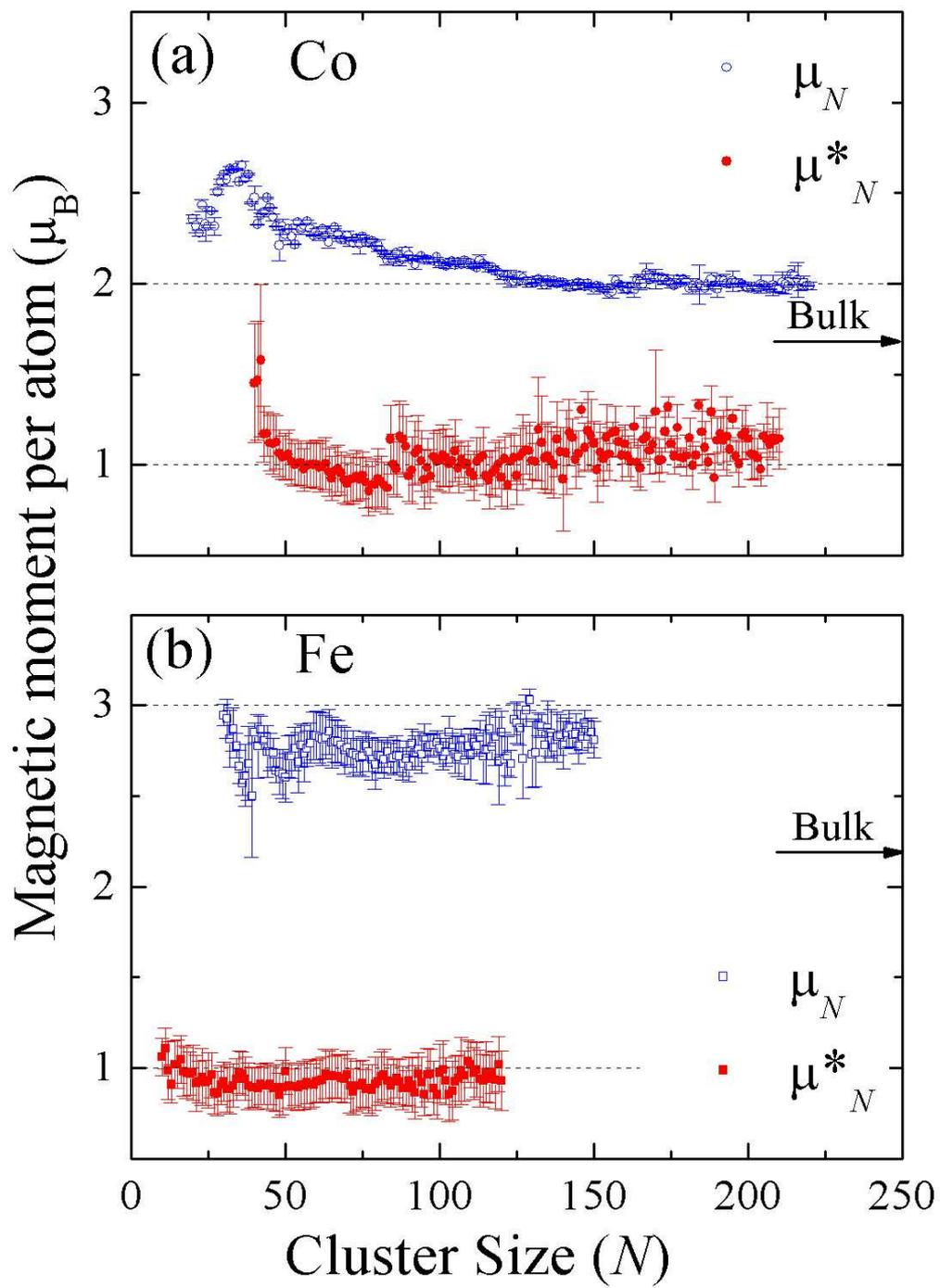

Figure 3

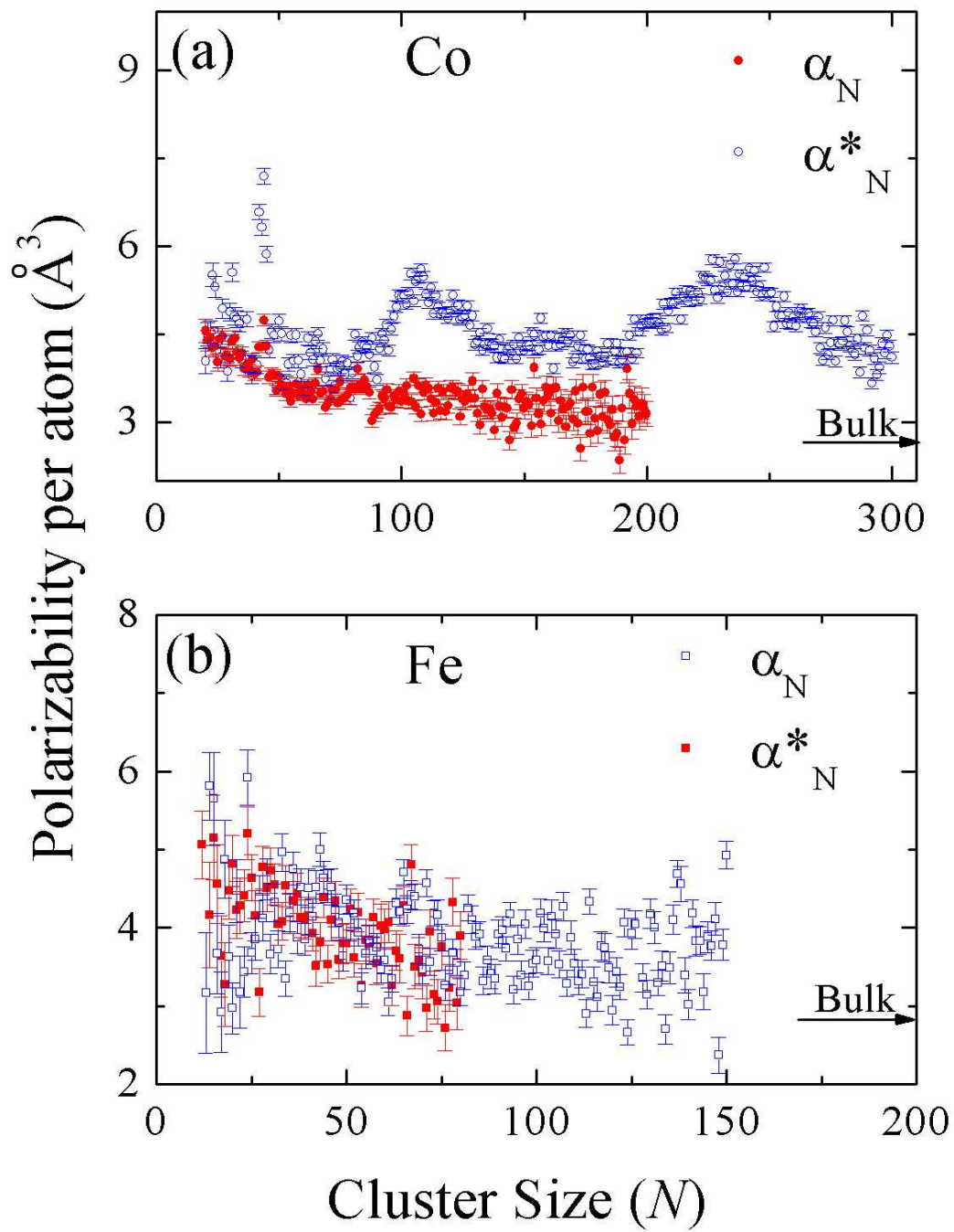

Figure 4

17